\documentclass[a4paper,11pt]{article}
\usepackage{pos}
\usepackage{physics}
\usepackage{graphicx}
\usepackage{subfig}

\title{Simulating Field Theories with Quantum Computers}

\author*[a]{Muhammad Asaduzzaman}
\author[b]{Simon Catterall}
\author[a]{Yannick Meurice}
\author[b]{Goksu Can Toga}

\affiliation[a]{University of Iowa,\\
 Department of Physics and Astronomy, Iowa City, IA 52246, USA.}

\affiliation[b]{Department of Physics, Syracuse University,\\
Syracuse, NY 13210, USA.}

\emailAdd{masaduzzaman@uiowa.edu}

\abstract{In this study, we investigate Trotter evolution in the Gross-Neveu and hyperbolic Ising models in two spacetime dimensions, using quantum computers. We identify different sources of errors prevalent in various quantum processing units and discuss challenges to scale up the size of the computation.  We present benchmark results obtained on a variety of platforms and employ a range of error mitigation techniques to address coherent and incoherent noise. By comparing these mitigated outcomes with exact diagonalization results and density matrix renormalization group calculations, we assess the effectiveness of our approaches. Moreover, we demonstrate the implementation of an out-of-time-ordered correlator (OTOC) protocol using IBM's quantum computers. }

\FullConference{The 40th International Symposium on Lattice Field Theory (Lattice 2023)\\
July 31st - August 4th, 2023\\
Fermi National Accelerator Laboratory\\}

\begin{document}
\maketitle

\section{Introduction}
Impressive advances in quantum information science (QIS) technologies in the last decade have instilled interest for many high-energy physicists to look into different physics problems where it might be possible to realize a \textit{quantum advantage}. Physicists focus on developing the field in many different directions. First and foremost, there are ongoing research efforts to identify problems that may benefit from the use of quantum computers. Second, there are endeavors to develop efficient quantum algorithms to gain quantum advantage. Thirdly, characterizing the present-day quantum devices and identifying technological limitations in developing more efficient algorithms to achieve the long-term goal of \textit{fault-tolerant quantum computation}. 

In this conference proceedings, we will discuss two different models that have importance in the field of high-energy physics. Firstly we discuss the Gross-Neveu model, a multi-flavor interacting fermionic model. We briefly discuss the qubitization of the model and demonstrate the real-time evolution of the system with quantum computers. The discussion is based on the published study of the authors \cite{asaduzzaman2022quantum} and is the earliest study that compares the implementation of a multi-flavor fermionic model on different quantum hardware platforms. 

Next, we discuss the real-time evolution of local observables for the Ising model on  hyperbolic space and compute the out-of-time ordered correlators (OTOC), which afford a measure of information spreading in a quantum system. For a detailed introduction to the model its classical simulation using density matrix renormalization group (DMRG) and time evolving block decimation (TEBD) please see the arXiv version of our study in \cite{asaduzzaman2023quantum}. The investigation into OTOCs and return probability serves as a benchmark result for determining the maximum system size feasible with present technologies. In both of our studies, we employ distinct error mitigation techniques that we have identified as crucial for obtaining compelling results with contemporary NISQ-era devices. These advancements in error mitigation bring us a step closer to addressing scattering problems with quantum devices on sizable lattices in the foreseeable future.

\section{Steps to analyze a QFT model with quantum devices \label{sec2}}
To compute the real-time evolution of the expectation values of a quantum field theory (QFT) model with quantum devices, there are well-established procedures that can be followed. 
\begin{itemize}
    \item First, a lattice description of the continuum Hamiltonian is derived.
    \item Next, the fields are converted into the qubit operators. Well-established techniques like the Jordan-Wigner transformation can be used to convert fermionic degrees of freedom to qubits. For gauge fields and scalar fields, usually, a cutoff is used to restrict the degrees of freedom to the region of phase space we are interested in. The choice of discrete levels in the fields is dubbed digitization in the literature.
    
    \item On the practical implementation side, the first step is to prepare the vacuum. Next, the initial state is prepared by applying the required gates for the wavepacket of interest.

    \item A time evolution circuit is developed by exponentiating the Hamiltonian derived in the second step. This boils down to applying single qubit rotation gates and different entangling operations between different sets of qubits.

    \item Finally, the measurement circuit is added after the time-evolution circuit.

    \item In the digital quantum devices these three steps: state preparation, time evolution of the state, and the measurement of the evolved state, are performed using a single quantum circuit. The projective measurement is performed many different times which we call `\textit{shots}'. Increasing the number of shots reduces the statistical error due to projective measurements.
\end{itemize}

There are usually two additional steps that need to be performed before implementing the quantum operations with the quantum devices.
\begin{itemize}
    \item Different operations of the quantum circuits are required to be expressed in terms of the basis gates of the device. Due to the technological differences, different platforms have different basis gates.
    
    \item If there is a restriction in the topology of the qubits in the quantum processing units, then there might be an efficient way to map the virtual qubits of our quantum field theory model to the physical qubits of the device. This is important since the bottleneck of the capacity of the current quantum devices is due to the lower fidelity of the entangling gates. 
\end{itemize}
Quantum devices are error-prone and we haven't yet achieved the fidelity requirement to implement quantum error correction in any of the existing technologies. Hence to improve the scope of the quantum simulation to obtain interesting results with the current NISQ-era devices, it is imperative to apply different \textit{`error mitigation'} techniques. Error mitigation techniques consist of characterizing different sources of coherent and incoherent noises, and devising ways to predict the expectation values of different operators in the absence of noise. Implementation of these techniques may introduce classical and quantum overheads.
\begin{itemize}
    \item \textbf{Readout Error Correction} To mitigate the error in the readout process of the measurement different error mitigation techniques can be applied: confusion matrix inversion, M3, Pauli-Twirling.

    \item \textbf{Incoherent noise mitigation} To mitigate incoherent noise zero-noise extrapolation can be used where measurements are performed scaling up the noise systematically and noisy simulation results are used to extrapolate the noiseless limit of the observables. There are also alternative approaches to mitigate incoherent noise like self-mitigation.

    \item \textbf{Coherent noise mitigation} To mitigate coherent noise, the typical strategy is to convert coherent noise into stochastic noise. Pauli Twirling can be used for this purpose. Another low overhead mititigation technique is dynamical decoupling which tackles decoherence of idle qubits.
\end{itemize}
    
\section{Gross-Neveu model}
\begin{figure*}[!htb]
\centering
	\subfloat{
		\label{schematic}
        \raisebox{0.5\height}{\includegraphics[width=.48\textwidth]{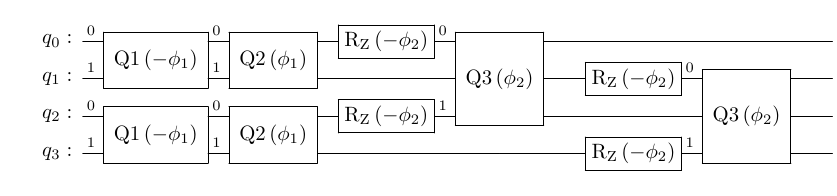}}%
	} \hfill
	\subfloat{
		\label{2flavor}
		\includegraphics[width=.48\textwidth]{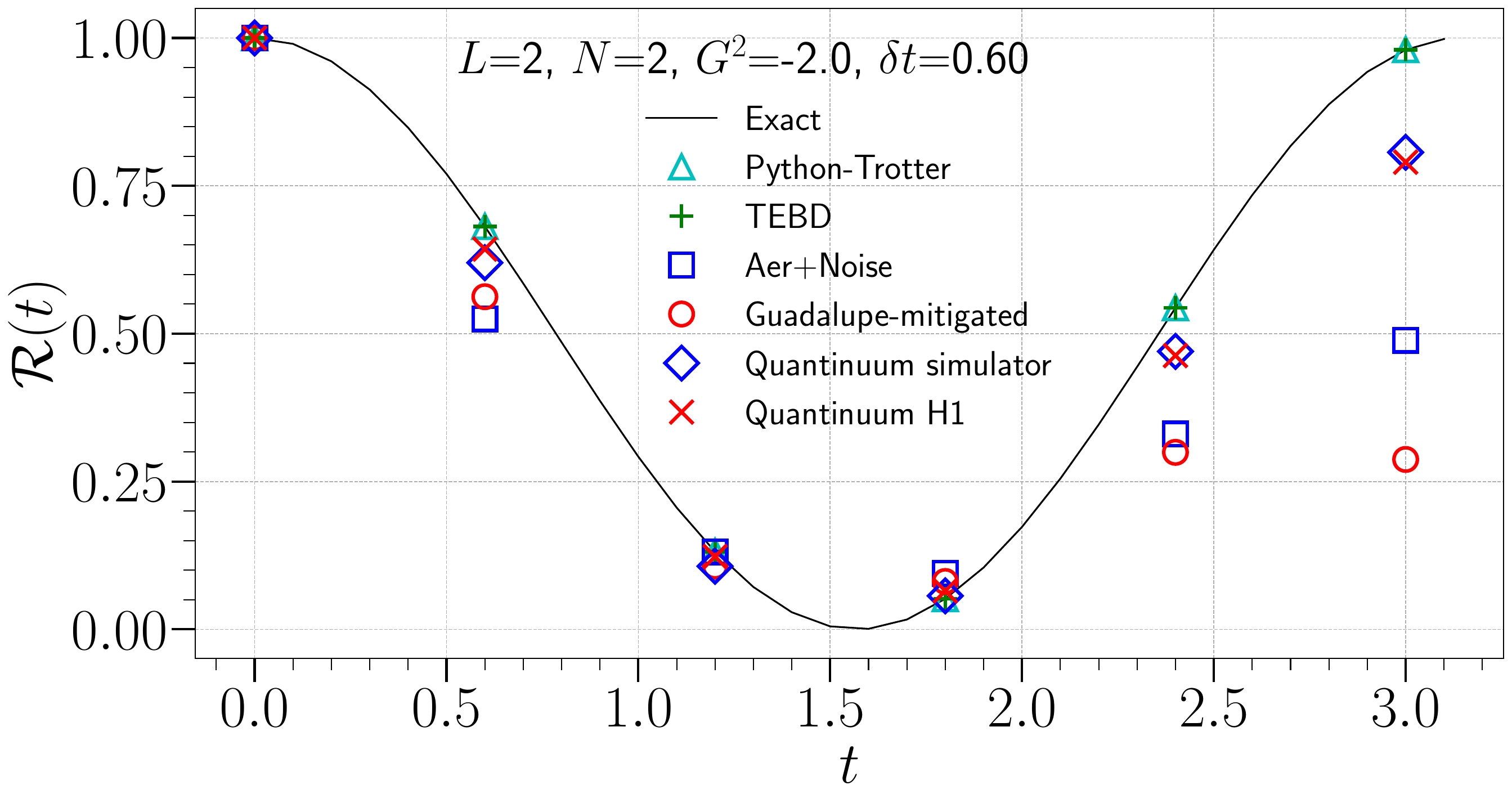}%
	}
	\caption{(a) Schematic diagram of the quantum operations in the circuit for a single step of Trotter evolution for
    two flavors. (b) Return probability $\mathcal{R}(t)$ is computed for an initial state $|\psi(0)\rangle=|0010\rangle$. The number of shots used for Guadalupe and Quantinuum devices are 4000 and 300 respectively. Adapted from \cite{asaduzzaman2022quantum}. }
\end{figure*}

The Gross-Neveu model \cite{Gross:1974jv} is a field theory of $N$ interacting Dirac fermions in (1+1) dimensions which has been the subject of many studies using
analytic and classical simulation methods.  In this section, we demonstrate the steps to compute real-time evolution which is closely related to performing scattering experiments with the quantum processing units. We can write down the lattice Hamiltonian for $L$-spatial sites and $N$-flavors of fermion in terms of reduced staggered fields $\chi^f_n$,
\begin{align}
    H^{(N)}=\sum_{n=1}^L\Bigg[i\sum_{f=1}^N\chi^{\dagger f}_n
    \left[\chi^f_{n+1}-\chi^f_{n-1}\right]+m\left(-1\right)^n\chi^{\dagger f}_n\chi^f_n + G^2
    \left(\sum_{f=1}^N\chi^{\dagger f}_n\chi^f_n\right)^2 \Bigg],
\end{align}
where, $f$ and $n$ denote the flavor and the site index respectively. In what follows
we consider the 2-flavor Gross-Neveu model. To investigate the model with quantum computers, we first require to convert the fermionic fields to bosonic Pauli operators. The \textit{qubitization} is performed using Jordan-Wigner transformation \cite{dargis_fermionization_1998}. For our study, we used open boundary conditions. The Hamiltonian can be written using Pauli-operators $\sigma \in \{X,Y,Z\}$ where the usual commutation relations of the Pauli operators hold,
\begin{align}
H^{(N)}&=  \sum_{n=1}^{L-1}  \Bigg[  \sum_{f=1}^{N} \Big( -X^f_n Y^f_{n+1}+Y^f_n X^f_{n-1} +\left(-1\right)^n m \, (1-Z^f_n ) +\frac{G^2}{2} \sum_{g,g>f} (I-Z^f_n) (I-Z^g_n)\Big) \Bigg].
\end{align}
We considered with staggered mass $m$ set to zero and used first order Trotter approximation for the real time ($t$) evolution of a state $\ket{\psi(t)}$ and computed the overlap with an initial state $\ket{\psi(0)}$. The return probability can be obtained from the time evolution operator $\mathcal{U}(t)=\exp(-i\,H^{(N)} t)$,
\begin{equation}
    \mathcal{R}(t)=\left|\langle \psi(0)|\mathcal{U}(t) | \psi(0) \rangle \right|^2.
\end{equation}
First order Trotter approximation allows us to approximate time-evolution  operator as products of two-qubit rotations ($Q_i$) and single qubit rotation $R_\sigma$ operators
\begin{align}
    \mathcal{U}(t)= \prod_{n,f,g:g>f} Q_1^f(n,n+1) Q_2^f(n,n+1) R_z^f(n) R_z^g(n) Q_3^{fg}(n,n).
\end{align}
To represent the qubitized $N$-flavor Gross-Neveu model with $L$-lattice sites, we require $Q=N \times L$ qubits. Figure~\ref{schematic}, demonstrates the mapping of the qubits and different quantum operations after the time evolution of one Trotter step $\delta t$. This whole block is repeated $n$ times to find the time evolution of an initial state after $t=n \delta t$ time.  The notation
$q_0$ and $q_1$ corresponds to degrees of freedom associated with one flavor of fermion, whereas $q_2$ and $q_3$ are associated with the second flavor of fermions. We can write down explicitly different entangling operations
\begin{align*}
    Q_1^f(n,n+1) &=\exp(-i \delta t X^f_n Y^f_{n+1} ) \to \exp(-i \delta t \, X_{fL+n} Y_{fL+n+1} ), \\
    Q_2^f(n,n+1) &=\exp(i \delta t Y^f_n X^f_{n-1} ) \to \exp(i \delta t \, Y_{fL+n} X_{fL+n-1} ), \\
    Q_3^{fg}(n,n)&=\exp(-i G^2 \delta t Z^f_n Z^g_{n} )  \to \exp(-i G^2 \delta t \,Z_{fL+n} Z_{gL+n} ).
\end{align*}
In the last step, we combine the flavor and the lattice-site index into a unique qubit index.

We computed the return probability using the IBM-Guadalupe quantum processing unit (currently retired) and the
Quantinuum's H1 series machine. We benchmark the results with $L=2$ sites. Results of the Trotter evolution are shown 
in Fig.~\ref{2flavor} for the $N=2$-flavor Gross-Neveu model. We prepare a low entangled matrix product state $\ket{\psi}=\ket{0100}$ and time evolve the state up to $n=5$ Trotter steps. Results obtained from the Quantum Processing Unit (QPU) are compared with different classical methods: exact diagonalization, Python-Trotter \footnote{Python-Trotter indicates implementation with a noiseless quantum device with first order Trotter approximation of the time evolution operator.}, and time-evolving block decimation (TEBD) techniques. We obsderve that there
is no significant difference between exact diagonalization and 
Python-Trotter results which validates the choice of the Trotter step for the computation.  

Simulator results incorporating the device noise model of the IBM platform demonstrate good agreement for small Trotter steps, whereas the largest Trotter step results did not match. However, to compare the results of the IBM device with the simulator we incorporated readout error mitigation.
On the other hand, the simulator result of the Quantinuum system
predicts results obtained on the Quantinuum machine quite well. 

Notably there is a significant difference in the results obtained from the Guadalupe and H1 machine. IBM's superconducting qubit architectures have restricted topology, resulting in the need for additional entangling (SWAP) gates, while Quantinuum's trapped ion machine has all-to-all connectivity. Hence, the Quantinuum machine does significantly better for larger Trotter steps. For example, the deviation for the 5\textsuperscript{th} Trotter step is $\sim 20 \%$ for the H1 machines and it $\sim 80\%$ for the Guadalupe machine. For a detailed discussion on the scaling of the system on lattice site $L$ and number of flavors $N$, ground state preparation with variational algorithms, and on the feasibility for solving scattering problems, readers are advised to consult the published paper \cite{asaduzzaman2022quantum}.

\section{Quantum Ising model on AdS space}
There are many interesting holographic connections between quantum field theories
and gravity on AdS spaces. The
quantum Ising model affords one of the simplest
toy models for investigating these issues using quantum
computers. We have hence studied the real time dynamics of
a system which consists of the transverse quantum Ising spin chain on
a discretization of one dimensional
hyperbolic space. The lattice Hamiltonian that describes the deformed Ising chain in (1+1) dimension can be represented as \cite{ueda_transverse_2010} 
\begin{align}
    H &= -J\sum_{<ij>}\frac{\eta_{i}+\eta_j }{2} S_i^z S_j^z -h\sum_i \eta_i S_i^x -m\sum_i \eta_i S_i^z.  \label{hamiltonian}
\end{align} 
Here, $S^p_i$ is a local spin operator at site $i$ with $p=\{x,y,z\}$. We considered odd numbers of lattice sites with an open boundary condition.
The deformation factors $ \eta_i=\cosh{\beta_i}$  can be thought of as square-roots of the metric determinant that appears due to the curvature of the underlying continuum
hyperbolic space. The constant curvature is associated with an intrinsic length scale $\ell_c$ and the local deformation factor is
\begin{equation}
\beta_i=\left(i-\frac{L-1}{2}\right)\frac{1}{\ell_c}.
\end{equation}
In the limit of $\ell_{c} \to \infty$, the planar transverse Ising model is recovered. 

\begin{figure*}[!htb]
\centering
	\subfloat{
		\label{szq_2d}
		\includegraphics[width=.48\textwidth]{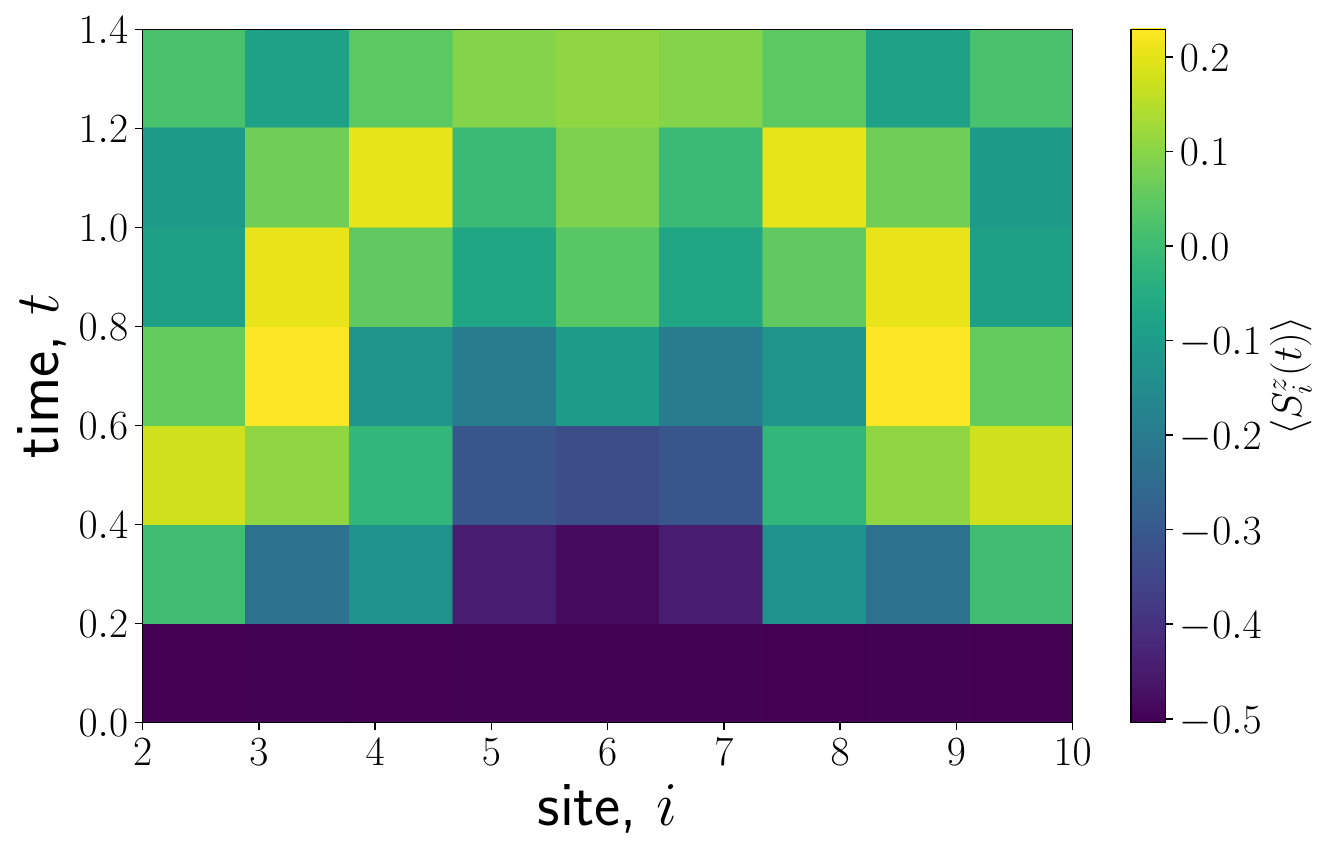}%
	} \hfill
	\subfloat{
		\label{mod_otoc_vary_Vqubit}
		\includegraphics[width=.44\textwidth]{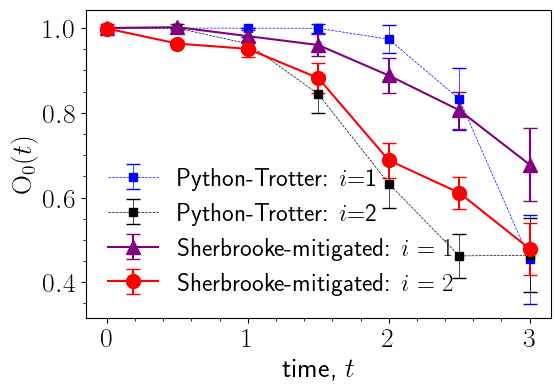}%
	}
	\caption{(a) Local magnetization $\langle S^z_i(t)\rangle$ computed using the QPU. Parameters: $N=13$, $J=2.0$, $h=1.05$, $\ell_{c}=2.0$. (b) Zeroth order modified OTOC as the position $i$ of the  $V$ operator varies. Parameters: $\ell_{c}=2.0,\,J=-0.5$, $h=-0.525$, $W(t)=\sigma^z_3(t)$, and $V=\sigma^z_i$. Adapted from \cite{asaduzzaman2023quantum}.}
\end{figure*}

Following steps similar to those outlined in section~\ref{sec2}, we have computed the local magnetization $\langle S^z_i(t)\rangle $ of a lattice chain with 13 spins. Fig~\ref{szq_2d} demonstrates time-evolution of the local magnetization computed using IBM's Guadalupe machine. We used dynamical decoupling, M3 readout error mitigation and zero-noise extrapolation for this calculation. For the details on the impact of the different error mitigation techniques, see the arXiv manuscript \cite{asaduzzaman2023quantum}.

We also computed modified out-of-time ordered correlators (OTOCs) using a randomized protocol developed by Vermersch et. al. \cite{vermersch2019probing} for a lattice chain of 7 sites. OTOCs (or out-of-time-ordered correlators) afford a method
to study the spread of quantum information under time evolution. In particular
they are useful observables to understand thermalization, the spread of
quantum information, and the onset of quantum chaos, in strongly interacting
quantum systems. 
Modified OTOCs of the $n$\textsuperscript{th} order can be computed using the following equation
\begin{equation}
    O_{n}(t)=\frac{\sum_{k_s \in E_n} c_{k_s} \overline{\langle W(t)\rangle_{u, k_s}\left\langle V^{\dagger} W(t) V\right\rangle_{u, k_0}}}{\sum_{k_s \in E_n} c_{k_s} \overline{\langle W(t)\rangle_{u, k_s}\langle W(t)\rangle_{u, k_0}}}.
\end{equation}
where, $W(t)$ and $V$ are local operators at different lattice sites. The expectation values are computed for many different random unitaries $u$, picked from the circular unitary ensemble (CUE). The parameter $k_s$ denotes different initial states. For a discussion of the different terms and it's connection to OTOC, see \cite{asaduzzaman2023quantum}.  The correlation of the expectation values $\langle W(t)\rangle$ and $\langle V^\dagger W(t) V \rangle$ gives a measure of the spreading of information over time.

The experiment was performed with IBM's Sherbrooke device. Fig.~\ref{mod_otoc_vary_Vqubit} clearly demonstrates that current NISQ era devices 
can already probe questions of information spreading and quantum chaos in simple
systems.
\section{Conclusion}
We have demonstrated how to implement real time evolution for two interesting
lattice quantum
field theories in (1+1) dimensions using different NISQ era quantum platforms
corresponding to superconducting qubits and trapped ions. The first of these - the Gross-Neveu model - describes systems of interacting relativistic fermions. The second - the
quantum Ising model in AdS space - is a toy model for understanding holography.
Our work illustrates several generic features and problems of current quantum
simulations.
 Firstly, our study highlights the present capabilities of quantum devices in handling non-local connectivity, a critical feature for applications in multi-flavor fermionic models and diverse field theoretic models such as the SYK model \footnote{For a recent study of the SYK model with quantum computers, consult \cite{Asaduzzaman:2023wtd}.}. Secondly, we underscore the effectiveness of contemporary error mitigation techniques. The use of the advanced mitigation techniques allowed us to simulate a 13-site lattice for a model with local interactions.  Thirdly, for a lattice with a 7 sites, we have demonstrated the computation of modified OTOCs, which bear the signature of quantum chaos. It would be an interesting endeavor to see if different characteristic times associated with the scrambling can be recovered with the current NISQ devices.

\section*{Acknowledgements}
We thank the IBM-Q hub at Brookhaven National Laboratory for providing access to the IBMQ quantum computers and Microsoft quantum credits program for the access of Quantinuum devices. S.C, M.A and G.T were supported under U.S.
Department of Energy grants DE-SC0009998 and DE-
SC0019139. Y. M. is supported under U.S. Department
of Energy grant DE-SC0019139.

\end{document}